\documentclass[runningheads]{llncs}
\usepackage{multirow}
\usepackage{graphicx}
\usepackage{amsfonts}
\usepackage{graphicx}
\usepackage{subfigure}
\usepackage{marvosym}
\begin{document}
\title{IREM: High-Resolution Magnetic Resonance Image Reconstruction via Implicit Neural Representation}

%
\titlerunning{HR MR image Reconstruction via Implicit Neural Representation}
%
\author{Qing Wu\textsuperscript{1}, Yuwei Li\textsuperscript{1}, Lan Xu\textsuperscript{1}, Ruiming Feng\textsuperscript{2}, Hongjiang Wei\textsuperscript{2}, Qing Yang\textsuperscript{3}, Boliang Yu\textsuperscript{1}, Xiaozhao Liu\textsuperscript{1}, Jingyi Yu\textsuperscript{1,(\Letter)}, and Yuyao Zhang\textsuperscript{1,4,(\Letter)\ \thanks{Yuyao Zhang and Jingyi Yu are co-corresponding author.}}}
\authorrunning{Q. Wu et al.}
%
\institute{\textsuperscript{1}School of Information Science and Technology, \\ ShanghaiTech University, Shanghai, China\\
\textsuperscript{2}School of Biomedical Engineering, Shanghai Jiao Tong University, Shanghai, China \\
\textsuperscript{3}Institute of Brain-Intelligence Technology, Zhangjiang Laboratory \\
\textsuperscript{4}Shanghai Engineering Research Center of Intelligent Vision and Imaging, ShanghaiTech University, Shanghai, China \\
\email{\{yujingyi, zhangyy8\}@shanghaitech.edu.cn}}
%
\maketitle              
\begin{abstract}
For collecting high-quality high-resolution (HR) MR image, we propose a novel image reconstruction network named IREM, which is trained on multiple low-resolution (LR) MR images and achieve an arbitrary up-sampling rate for HR image reconstruction. In this work, we suppose the desired HR image as an implicit continuous function of the 3D image spatial coordinate, and the thick-slice LR images as several sparse discrete samplings of this function. Then the super-resolution (SR) task is to learn the continuous volumetric function from a limited observations using a fully-connected neural network combined with Fourier feature positional encoding. By simply minimizing the error between the network prediction and the acquired LR image intensity across each  imaging plane, IREM is trained to represent a continuous model of the observed tissue anatomy. Experimental results indicate that IREM succeeds in representing high frequency image feature, and in real scene data collection, IREM reduces scan time and achieves high-quality high-resolution MR imaging in terms of SNR and local image detail.

\end{abstract}
\section{Introduction}
\par High Resolution (HR) medical images provide rich structural details to facilitate early and accurate diagnosis. In MRI, unpredictable patient movements result in difficulties to collect artifact-free high-resolution 3D images at sub-millimeter within a single scan. Thus, multiple anisotropic low-resolution (LR) MR scans with thick slice are always conducted at different scan orientations with a relatively short scan time to eliminate motion artifacts. MRI Super-Resolution (SR) techniques are then used to enhance the image spatial resolution from one or more LR images. According to the number of processed images, related works can be divided into two categories: multiple image super-resolution (MISR) \cite{Shi,Chen2018} or single image super-resolution (SISR) \cite{Rueda}.
\par MRI SR techniques were originally proposed as MISR algorithms \cite{Peled}. Multiple thick-sliced LR images of the same subject were acquired to reconstruct an HR image. Gholipour et al. \cite{Gholipour} developed a model-based MISR technique (always known as SRR) that enabled isotropic volumetric image reconstruction from arbitrarily oriented slice acquisitions. SRR achieved great success and was broadly applied for motion-prone MR acquisitions (i.e., fetus brain construction \cite{Michael}). Jia et al. \cite{Jia} proposed a Sparse Representation based SR approach using only the multiple LR scans to build the over-complete dictionary. Previous studies \cite{Gholipour,Rahman,Reeth,Yang} showed that for MISR methods, in a given acquisition time, when the alignment among multiple frames is known or correctly estimated, the SR reconstructed image presented higher SNR (thick-slice improves SNR in single voxel) when comparing with the isotropic image acquired with similar spatial resolution. But it is challenging for MISR methods to build submillimeter-level image detail, such as in brain cerebellum.
\par On the other hand, SISR is attracting attentions due to its advantage of shorter scan time to further reduce the motion artifacts. Recently, the performance of SISR methods were significantly improved by Deep convolution neural network (CNN) because of its non-linearity to imitate transformation between LR and HR images. Various network architectures for MRI SISR have been proposed \cite{lyu2020multi,lyu2020mri,delannoy2020segsrgan,pham2017brain,chen2020mri}. However, there have to exist a large HR image dataset as ground truth to train the SR network and the learned model lacks generalization for the input images with different qualities, e.g., different spatial resolutions, SNRs. As a result, this kind of methods cannot handle SR tasks when no HR ground truth is available (i.e., fetus brain reconstruction).
\par In this paper, we propose IREM, which is short for \itshape{Implicit REpresentation for MRI}. \upshape Our work intends to further improve MISR reconstruction performance via deep neural network. Instead of CNN architecture, we train a fully-connected deep network (MLP) to learn and represent the desired HR image. Specifically, we suppose the HR image as an implicit continuous function of the 3D image spatial coordinate, and the 2D thick-slice LR images as several sparse discrete samplings of this function. Then the SR task is to learn the continuous volumetric function from a limited 2D observations using an extended MLP. The MLP takes the positional encoded 3D coordinate $\left(x,y,z\right)$ as input and outputs the image intensity $I\left(x,y,z\right)$ of the voxel at that location. The positional encoding is condected via Fourier feature mapping \cite{tancik2020fourier} to project the input 3D coordinates into a higher dimensional space, for enabling the network to learn sufficient high frequency image features from the LR image stacks. By minimizing the error between the network prediction and the acquired LR image intensity across each LR imaging plane, IREM is trained to reconstruct a continuous model of the observed tissue anatomy. We demonstrate that our resulting HR MRI reconstruction method quantitatively and qualitatively outperforms state-of-the-art MISR methods. The major advantage of IREM is summarized as below:
\begin{itemize}
    \item No need for large amount of HR data set for training the network; 
    \item No constraints for input LR image resolution and up-sampling scale; 
    \item Comparing with the realistic HR image acquisition with similar total scan time, the SR image provides equivalent level of image quality and higher SNR.
\end{itemize}
\section{Method}
\par Inspired by a very recent implicit scene representation work \cite{Mildenhall}, we build a deep learning model that learn to represent an image function $I(x,y,z)$ from input voxel location $(x,y,z)$. In model training, the spatial coordinate of the LR images are input into a fully-connected network with positional encoding, while the model is trained to output the image intensity $I(x,y,z)$ of the voxel at that location with minimizing error comparing with the LR image intensity. Then, the fine-trained network reconstruct the desired HR image with arbitrary up-sampling SR rate (with arbitrary input $(x,y,z)$). The spatial locations that are not observed in the LR images are thus the “test data” generated by the proposed SR model.
\begin{figure}[htb]
\centering
\includegraphics[width=\textwidth]{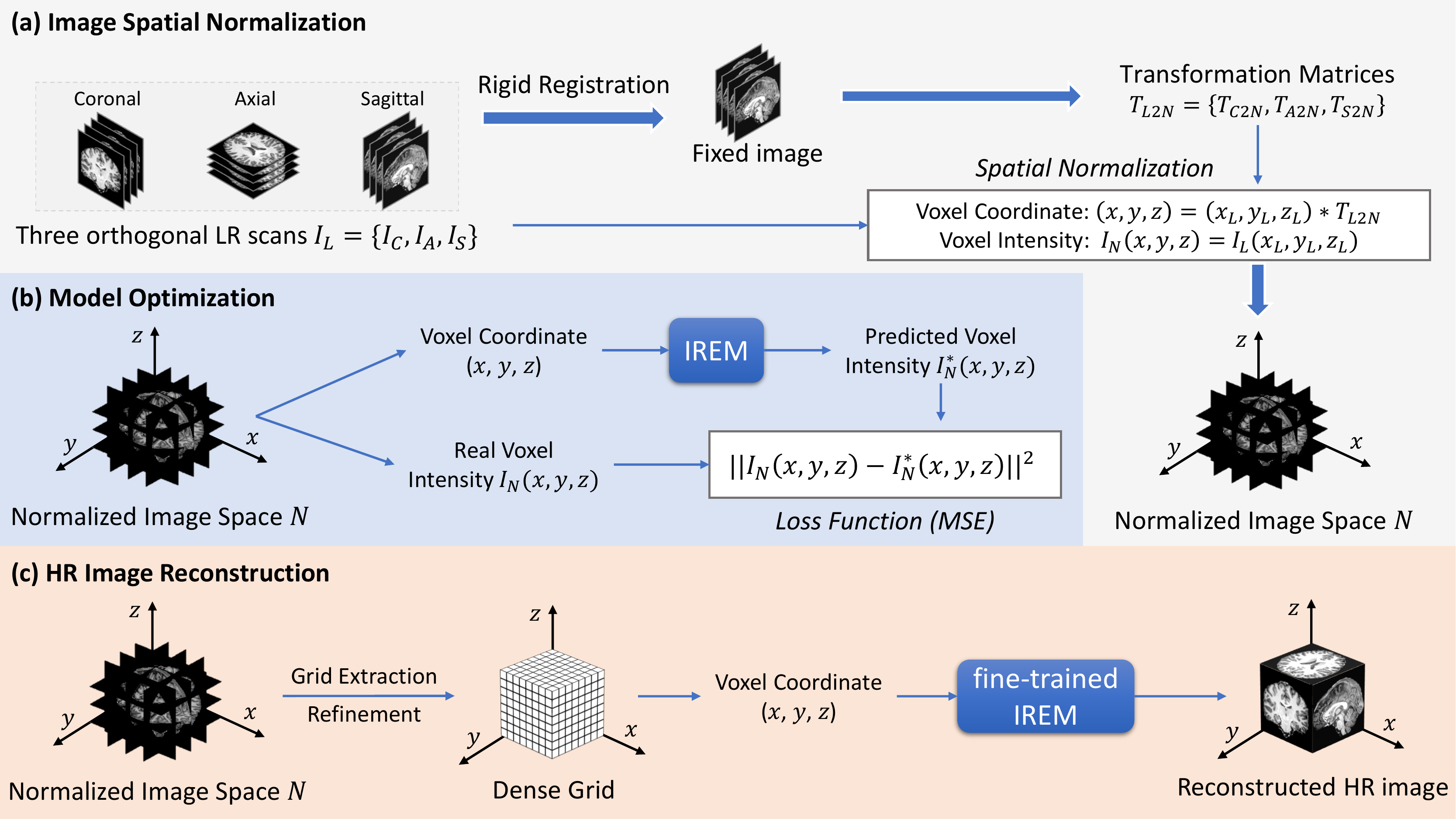}
\caption{An overview of the proposed IREM approach.} 
\label{fig1}
\end{figure}
\par An overview of IREM is depicted in Figure \ref{fig1}, which is presented in three stages: a) Image Spatial Normalization; b) Model Optimization; c) HR Image Reconstruction. 
\subsection{Image Spatial Normalization}
\par As demonstrated in Figure 1 (a), we first build a set of anisotropic LR (thick-sliced MR scanning) image stacks $I_L=\{{I}_C,I_A,I_S\}$. For simplifying, we use three orthogonal scanning orientations (coronal, axial, sagittal) to represent the LR stacks. Then we randomly select one from the three LR scans as fixed image and rigidly register all the LR scans to the fixed image to generate transformation matrices $T_{L2N}=\{T_{C2N},T_{A2N},T_{S2N}\}$. Finally, we utilize transformation matrices $T_{L2N}$ to transfer each LR image from its original space to the normalized image space $N$.
\par In the normalized space, the image intensities from different LR image stacks at same coordinate $(x,y,z)$ represents coherent observations of the same tissue anatomy in different imaging orientations. Note that the pair of voxel coordinate and intensity in the normalized space $N$: $\left(x,y,z\right){\rightarrow}I_N\left(x,y,z\right)$ are transformed from LR scans at position $\left(x_L,y_L,z_L\right)\rightarrow I_L\left(x_L,y_L,z_L\right)$ by transformation matrix (i.e., $\left(x,y,z\right)=\left(x_L,y_L,z_L\right)\ast T_{L2N}$ and $I_N\left(x,y,z\right)=I_L\left(x_L,y_L,z_L\right))$.
\subsection{Model Optimization} 
\par The process of model optimization is illustrated in Figure 1 (b). After the image spatial normalization, the training data set for building IREM model is prepared. We feed the voxel coordinate $\left(x,y,z\right)$ into IREM to compute the predicted voxel intensity $I_N^\ast\left(x,y,z\right)$. Then, we optimize IREM by minimizing the mean square error loss function between the predicted voxel intensity and the real observed voxel intensity across the LR imaging planes. The loss function $L$ is denoted as:
\begin{equation}
\label{equ-loss}
L(\theta)=\frac{1}{\mathcal{K}} \sum_{(x, y, z) \in N}\left\|I_{N}(x, y, z)-I_{N}^{*}(x, y, z)\right\|^{2}
\end{equation}
where $\mathcal{K}$ is mini-batch size, $N$ is normalized image space, and $\theta$ denotes IREM’s parameters.
\subsubsection{Architecture of IREM.}As illustrated in Figure 2, IREM consists of a position encoding section (via Fourier feature mapping \cite{tancik2020fourier}), and a fully-connected network (MLP). It takes 3D spatial coordinate as input and outputs the intensity of the voxel at that location.
\begin{figure}
\centering
\includegraphics[width=\textwidth]{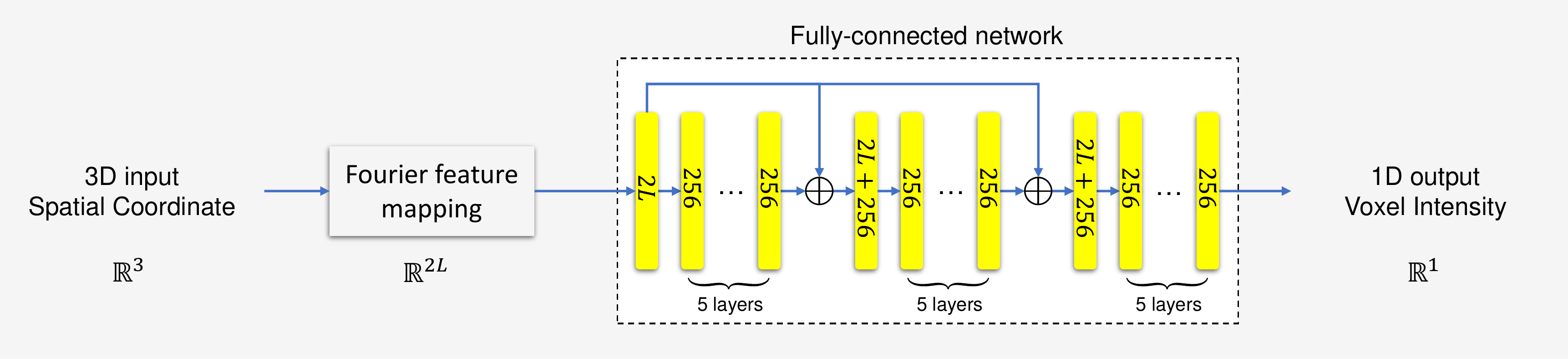}
\caption{Architecture of IREM.} 
\label{fig2}
\end{figure}
\paragraph{Fourier feature mapping.} Theoretically, a simple multi-layer perceptron can approximate any complicated functions \cite{Hornik}. Recently, Rahaman et al. \cite{rahaman2019spectral} found that deep learning networks are biased towards learning lower frequency functions in practical training. To this end, Mildenhall et al. proposed Fourier feature mapping \cite{tancik2020fourier} that maps the low-dimensional input to a higher dimensional space and thus enable network to learn higher frequency image feature. In IREM, we perfrom Fourier feature mapping \cite{tancik2020fourier} to map the 3D voxel coordinates to a higher dimensional space $\mathbb{R}^{2L} (2L\ >3)$ before passing them to the fully-connected network. Let $\gamma\left(\cdot\right)$ denotes Fourier feature mapping from the space $\mathbb{R}^3$ to $\mathbb{R}^{2L}$ and it is calculated by
\begin{equation}
    \gamma\left(\mathcal{P}\right)=\left[\cos{\left(2\pi \mathcal{BP}\right)},\sin{\left(2\pi \mathcal{BP}\right)}\right]^T
\end{equation}
where $\mathcal{P}=\left(x,y,z\right)\in \mathbb{R}^3$ and each element in $\mathcal{B}\in \mathbb{R}^{L\times3}$ is sampled from gaussian distribution $\mathcal{N}\left(0,1\right)$. 
\paragraph{Fully-connected network.} The network has eighteen fully-connected layers. Each fully-connected layer is followed by a batch normalization layer \cite{Ioffe} and a ReLU activation \cite{Kaiming}. In order to ease the difficulty of optimizing IREM, we include two skip connections that concatenate the input of the fully-connected network to the 6th layer’s activation and the 12th layer’s activation, respectively. And except that there are $2L$, $2L+256$, $2L+256$ neurons in the 1st, 7th, and 13th fully-connected layers. Other fully-connected layers have all 256 neurons.
\subsection{HR Image Reconstruction}
\par The sketch map of an HR image reconstruction by a fine-trained IREM is shown in Figure 1 (c). We build a dense grid (i.e., a coordinate set with more voxels than in LR stacks) in the normalized image space $N$, with isotropic image resolution. Then, we pass each voxel coordinate $\left(x,y,z\right)$ in the grid into IREM to produce predicted the voxel intensity $I_N^\ast\left(x,y,z\right)$. An HR image thus is reconstructed.

\section{Experiments}
\subsection{Data}
\par We conduct experiments on three datasets. The details are shown in Table \ref{tab1}. Dataset \#A consists of T1-weighted (T1w) HR brain MR scans from five healthy adults on a 7T MR scanner and \#B consists of 3T T2 flair images from two patients with lesions in brain white matter. Dataset \#C consists of four T1w brain MR scans (An HR scan used for reference and three orthogonal LR scans used for input images) from the same volunteer. 
\par \textbf{Evaluation for the accuracy of implicit HR image representation on simulation data.} For dataset \#A and \#B, we employ the original MR image as ground truth (GT) HR image and downsample the image by the factor of $\{4,\ 8\}$ in the three image dimensions, to simulate three orthogonal LR scans. We then use the built LR stacks to train our model, and compare the reconstructed HR image with the GT to evaluate the ability of IREM to learn the high frequency image content in the implicit HR image. 
\par \textbf{Evaluation for the performance in real data collection protocol.} We conduct a real data collection scene using IREM on dataset \#C. In this case, no actual GT can be built. We scan a HR T1w HR brain image with $0.7$×$0.7$×$0.7$ $mm^3$ isotropic in 30.9 mins as HR reference. Then three anisotropic LR images are scanned in coronal, axial, and sagittal orientations in about 10 mins each, so the total scan time of 3 LR stacks is about 2 mins shorter than the HR scan. 
\begin{table}[t]
\centering
\caption{Details of the MR image datasets used in the experiments.}\label{tab1}
\scalebox{0.9}{
\begin{tabular}{|c|c|c|c|c|c|}
\hline
Dataset Name         & \# Image      & Number of Slices & Matrix Size & Voxel Size ($mm^3$) & Scan Time \\ \hline
\#A                  & sub1-sub5     & 244              & 244×244     & 0.8×0.8×0.8 & -         \\ \hline
\#B                  & sub1-sub2     & 64               & 256×256     & 1×1×2       & -         \\ \hline
\multirow{4}{*}{\#C} & HR reference  & 254              & 368×345     & 0.7×0.7×0.7 & 30.9 mins      \\ 
                     & axial scan    & 80               & 320×320     & 0.7×0.7×2.8 & 10.3 mins       \\ 
                     & coronal scan  & 80               & 320×320     & 0.7×2.8×0.7 & 10.3 mins       \\  
                     & sagittal scan & 64               & 320×320     & 2.8×0.7×0.7 & 8.2 mins       \\ \hline
\end{tabular}}
\end{table}
\subsection{Implementation details.}
\par We adopt Adam optimizer \cite{Kingma} to train IREM through back-propagation with a mini-batch size of 2500, and the hyperparameters of the Adam are set as follows: $\beta_1=0.9,\ \beta_2=0.999,\ \varepsilon={10}^{-8}$. The learning rate starts from ${10}^{-4}$ and decays by factor 0.5 every 500 epochs. For fair comparison, we implement two MISR methods, including Super-Resolution Reconstruction \cite{Michael} (SRR) and B-spline interpolation, to compare with proposed work. We assess quantitatively the performance of the three methods in terms of peak signal-to-noise ratio (PSNR), structural similarity (SSIM) \cite{Zhou}. 
\subsection{Results}
\begin{figure}
    \centering
    \includegraphics[width=0.9\textwidth]{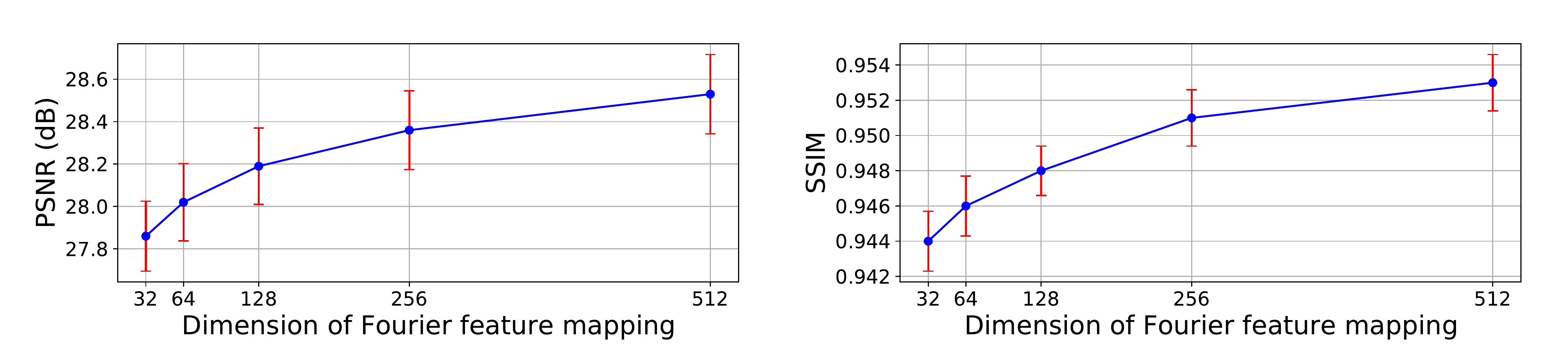}
    \caption{Evolution of the performance of IREM on the different dimensions $2L$ of Fourier feature mapping \cite{tancik2020fourier} for dataset \#A. Here the factor $k$ of down-sampling is 8.}
    \label{fig3}
\end{figure}
\subsubsection{Effect of the Fourier feature mapping.} Based on dataset \#A, we investigate the effect of Fourier feature mapping \cite{tancik2020fourier} on the performance of IREM. As shown in Figure \ref{fig3}, the performance of IREM improves with the dimension $2L$ of Fourier feature mapping \cite{tancik2020fourier}, but its growth rate gradually decreases, which is consistent with the previous work \cite{tancik2020fourier}. To achieve the balance between efficiency and accuracy, the dimension $2L$ is set as 256 in all the experiments below.
\begin{figure}[htb]
    \subfigure[On dataset \#A.]{
    \begin{minipage}[t]{0.485\linewidth}
    \centering
    \includegraphics[width=1\textwidth]{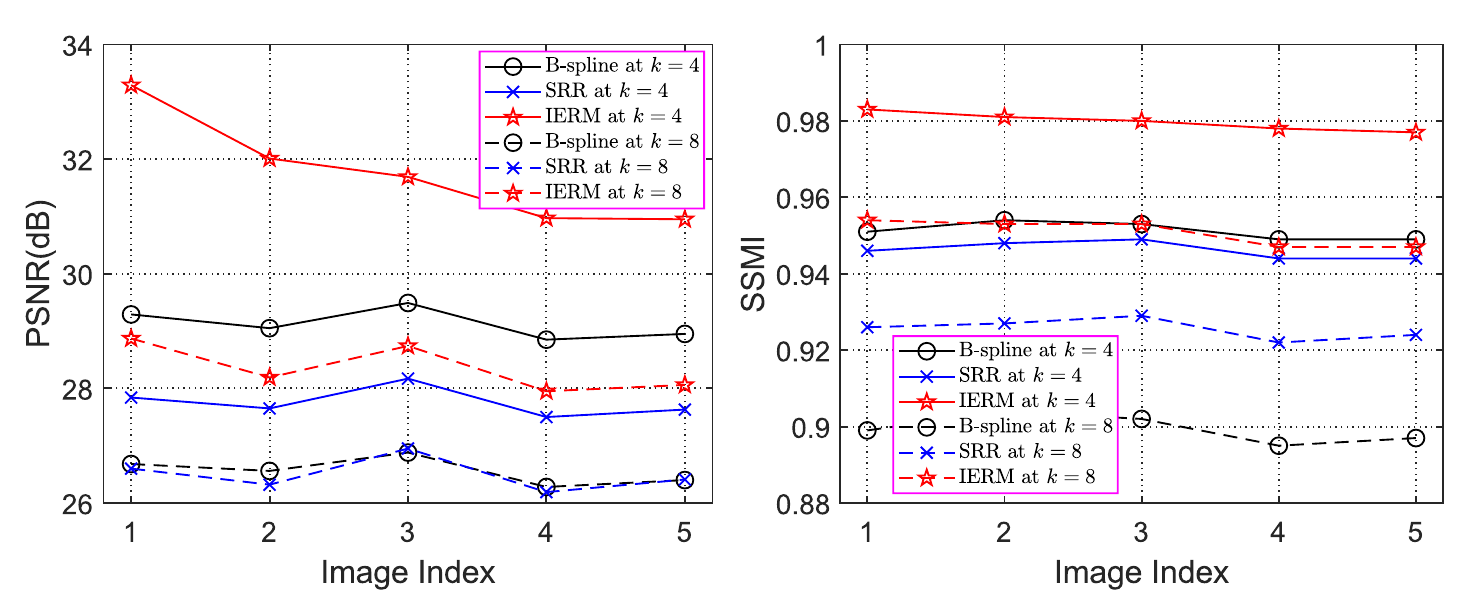}
    \end{minipage}}
    \subfigure[On dataset \#B.]{
    \begin{minipage}[t]{0.485\linewidth}
    \centering
    \includegraphics[width=1\textwidth]{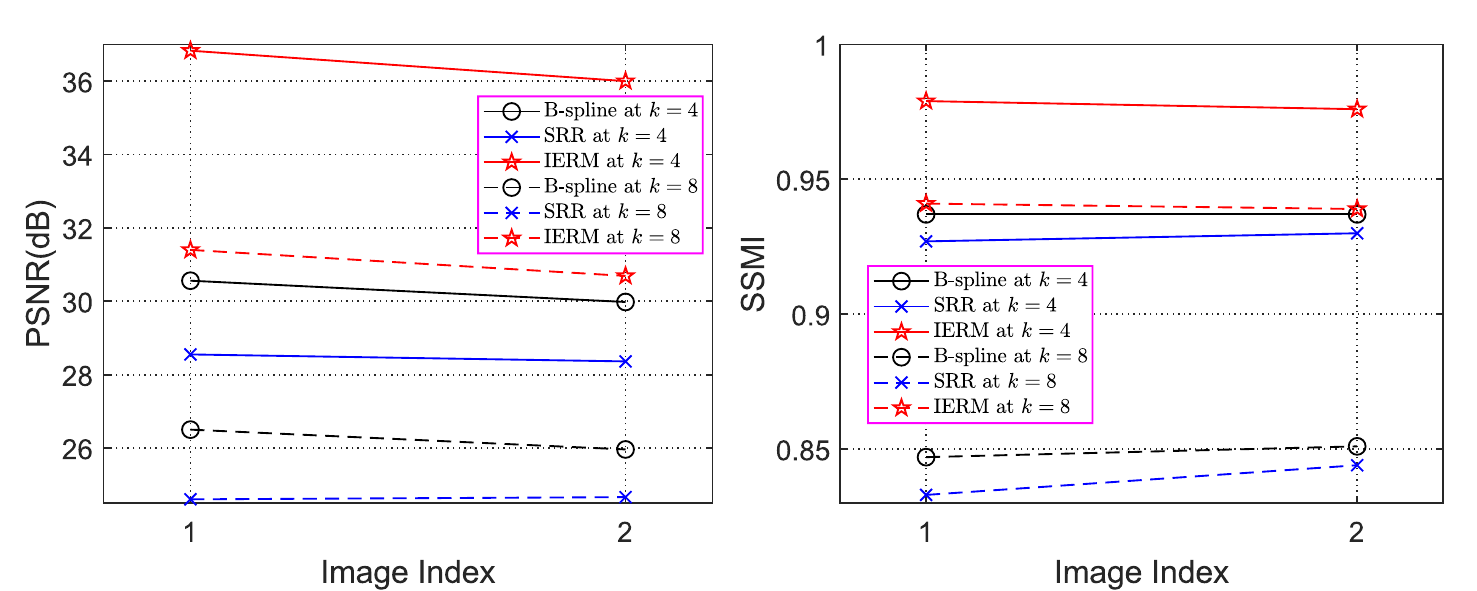}
    \end{minipage}}
    \centering
    \caption{Quantitative results (PSNR(dB) / SSMI) of B-spline interpolation, SRR \cite{Michael}, and IREM on dataset \#A and \#B. Here $k$ denotes the factor of down-sampling.}
    \label{fig7}
\end{figure}
\begin{figure}
    \centering
    \includegraphics[width=\textwidth]{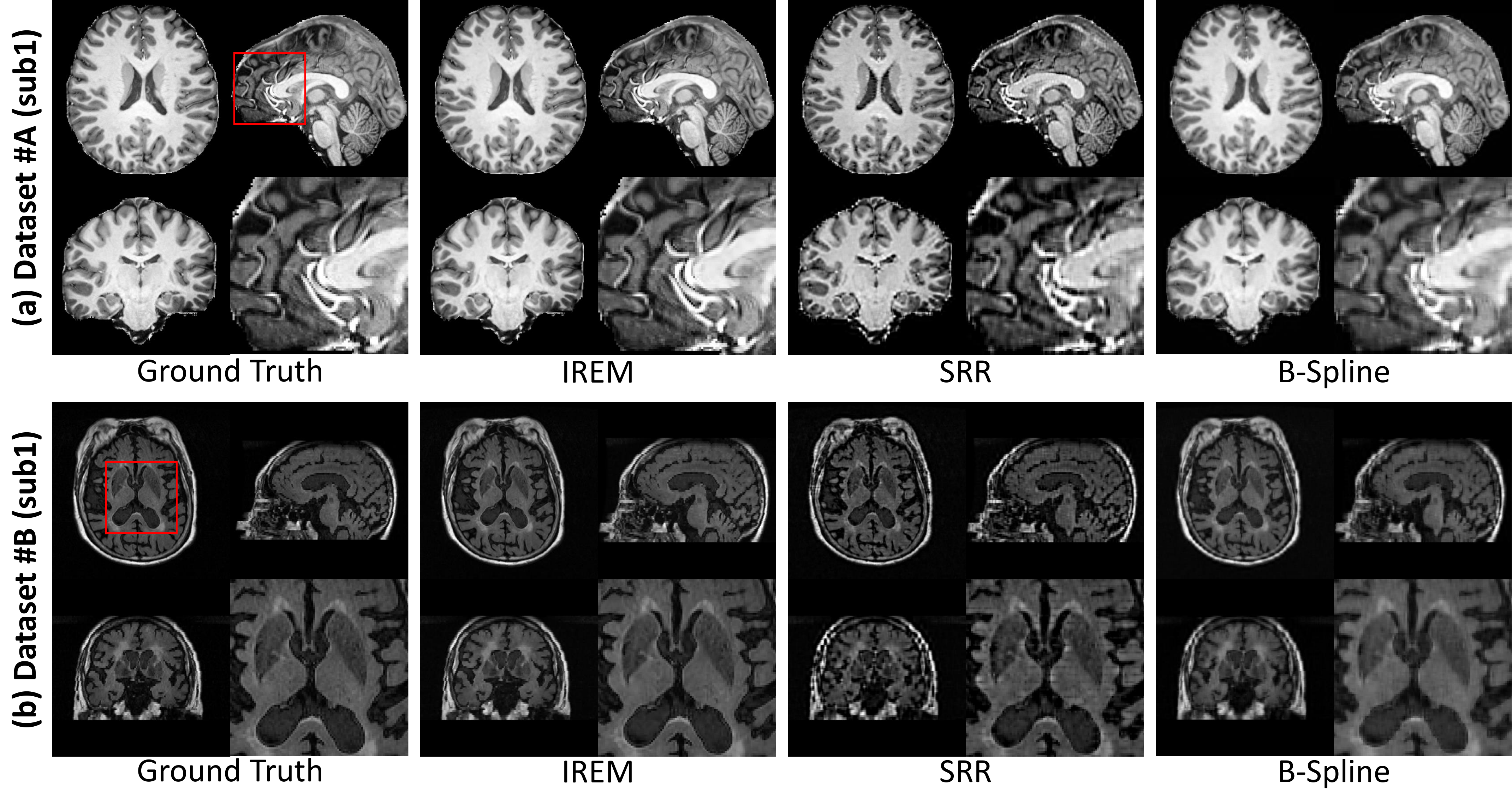}
    \caption{Qualitative results of B-spline interpolation, SRR \cite{Michael}, and IREM on dataset \#A (sub1) and \#B (sub1). Here the factor $k$ of down-sampling is 4.}
    \label{fig5}
\end{figure}
\subsubsection{Performance on simulation validation.} On dataset \#A, the quantitative evaluation results of B-spline interpolation, SRR \cite{Michael}, and IREM are shown in Figure \ref{fig7} (a). We indicate that IREM consistently outperforms the two baselines at all factors $k$ of down-sampling in terms of all metrics. Figure \ref{fig5} (a) demonstrates that the representative slices of an HR image (\#A sub1) reconstructed from the three methods, and the results from IREM visually are closest to GT HR image.
\par On dataset \#B, the quantitative and qualitative evaluation results of the three methods are shown in Figure \ref{fig7} (b) and Figure \ref{fig5} (b), respectively. We can see that IREM achieves the best performance, which is consistent with the experiments on dataset \#A.
\begin{figure}
    \centering
    \includegraphics[width=\textwidth]{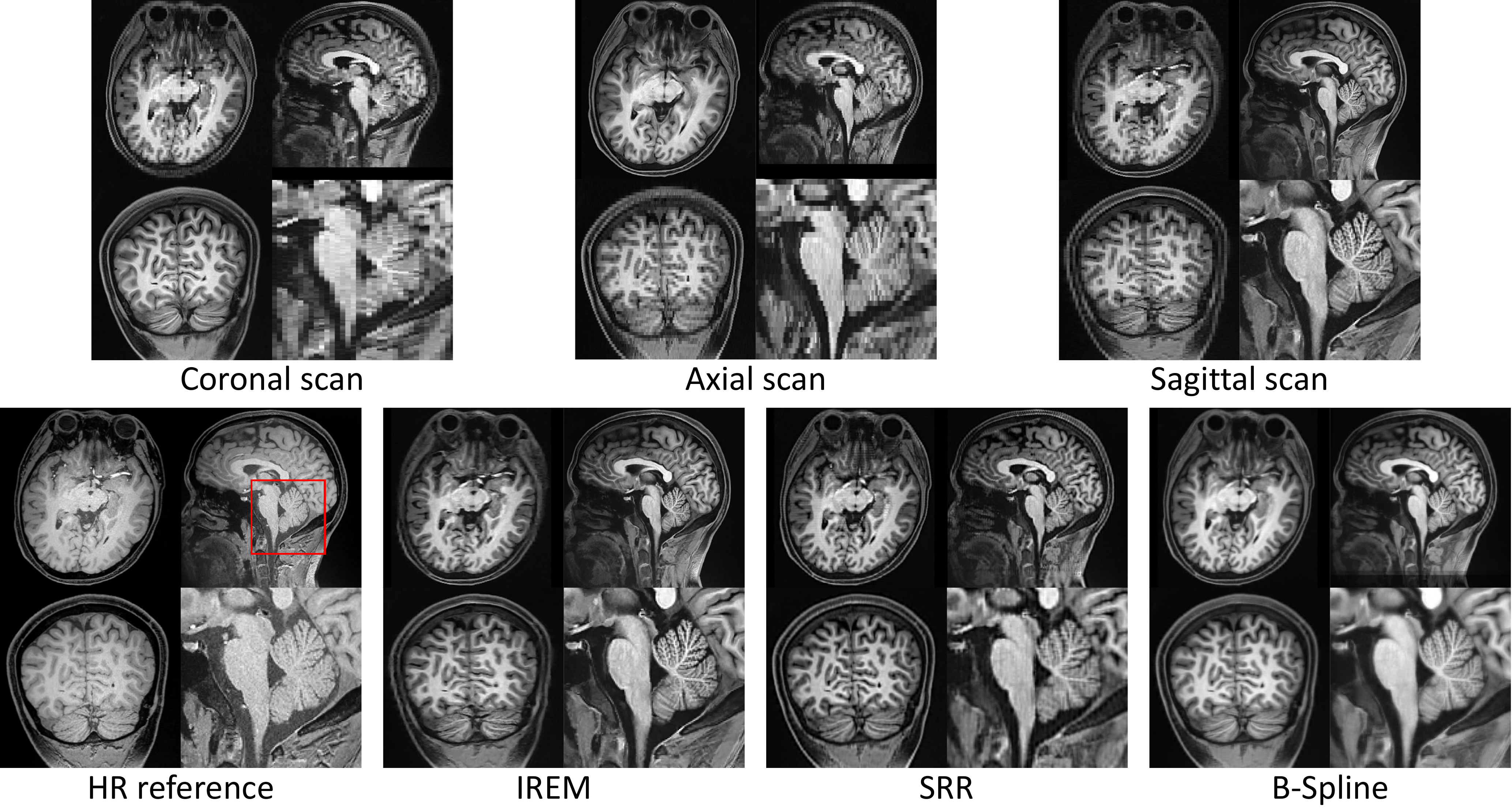}
    \caption{Results of B-spline interpolation, SRR \cite{Michael}, and IREM on dataset \#C.}
    \label{fig6}
\end{figure}
\subsubsection{Performance in real data collection protocol.} Figure \ref{fig6} shows the results of the three methods on dataset \#C. All MISR algorithms are conducted after a unique image spatial normalization step. Comparing with IREM, the image built from B-Spline is more blurry and SRR \cite{Michael} yields an image with artifacts. As indicated in the enlarged part, the image built from IREM achieves equivalent qualitative image details comparing with the HR reference in the cerebellum, which is one of the most complicated anatomy in human brain. Besides, benefitting from the multiple anisotropic thick slice scanning strategy, the SNR in each voxel of the LR image stack is about 16 times (definition and computational detail of SNR can be found in \cite{scherrer2012super}) higher than that in the reference HR image. Thus the image contrast between white matter and gray matter in the reconstructed image is better than that in the reference HR image. The result suggests that IREM is a more effective and reasonable pipeline to achieve high-quality HR image comparing with scanning directly an isotropic HR image.
\section{Conclusion}
\par In this paper, we proposed IREM, a novel implicit representation based deep learning framework to improve the performance of multiple image super-resolution task. In IREM, we combined spatial encoding and fully connected neural network, then trained a powerful model that precisely predict MR image intensity from input spatial coordinates. The HR reconstruction result on two set of simulated image data indicated the ability of IREM to accurately approach the implicit HR space. While the real scene data collection demonstrated that based on multiple low-resolution MR scans with thicker slice, IREM reconstructed HR image with both improved anatomy detail and image contrast.
\section{Acknowledgements}
\par This study is supported by the National Natural Science Foundation of China (No. 62071299, 61901256).
\bibliographystyle{splncs04}
\bibliography{refs}
\end{document}